\documentclass[12pt]{revtex4}
\usepackage{dcolumn}
\usepackage{graphicx}
\usepackage{amsmath}
\usepackage{amsfonts}
\usepackage{amssymb}
\usepackage{psfrag}
\usepackage{wrapfig}
\usepackage{subfigure}
\usepackage{makeidx}
\usepackage{bm}
\usepackage{epsf}
\usepackage{hyperref}

\begin{document}

\title{A new metric for rotating black holes in Gauss-Bonnet gravity}

\author{ Rui-Hong Yue$^{1}$\footnote{Email:yueruihong@nbu.edu.cn},
De-Cheng Zou$^{2}$\footnote{ Email:zoudecheng789@163.com},Tian-Yi Yu$^{1}$
 and Zhan-Ying Yang$^{2}$\footnote{ Email:zyyang@nwu.edu.cn}}

\affiliation{ $^{1}$Faculty of Science, Ningbo University, Ningbo, 315211, China\\
$^{2}$Department of Physics, Northwest University, Xi'an, 710069, China}
\date{\today}

\begin{abstract}
\indent

This paper presents a new metric and studies slowly rotating Gauss-Bonnet black holes
with one nonvanishing angular momentum in five dimensional anti-de Sitter spaces.
Taking the angular momentum parameter $a$ up to second order, the slowly rotating
black hole solutions are obtained by working directly in the action. In addition,
it also finds that this method is applicable in higher order Lovelock gravity.
\end{abstract}



\maketitle

\section{Introduction}  
\label{11s}
It is a general belief that Einstein's gravity is low-energy limit of a quantum theory
of gravity. Lovelock \cite{Lovelock:1971yv} extended the Einstein tensor, which
is only symmetric and conserved tensor depending on the metric, to the most general tensor.
In higher dimensional spacetimes, the Lovelock theory is the most nature extension of
general relativity and its field equations of motion contain the
most symmetric conserved tensor with no more than two derivative of the
metric. It has been argued that the Gauss-Bonnet term appears as the leading correction to
the effective low energy action of the string theory. Until now, the analytic expressions
of static and spherically symmetric Gauss-Bonnet black hole solutions have been investigated
in \cite{Boulware:1985wk, Wheeler:1985qd, Wheeler:1985nh}, and of Born-Infeld-Gauss-Bonnet
models in \cite{Wiltshire:1988uq, Wiltshire:1988jq, Aiello:2004rz}.
The thermodynamics of the uncharged static spherically Gauss-Bonnet black hole solutions
have been considered in \cite{Cai:2003gr, Cai:2001dz, Cho:2002hq} and of charged
solutions in \cite{Cvetic:2001bk}.
R. A. Konoplya et al. \cite{Konoplya:2004xx, Abdalla:2005hu}
presented an analysis of the scalar perturbations in the background
of Gauss-Bonnet black hole spacetimes and its (in)stability in high dimensions.
Very recently the quasinormal mode of a scalar field in five-dimensional Lovelock black
hole spacetime for different angular quantum numbers $l$ has been
obtained in \cite{Chen:2009an}. Liu \cite{Liu:2007zze} studied the electromagnetic
perturbations of black holes in Gauss-Bonnet gravity.

For Gauss-Bonnet gravity, it is interesting to explore some rotating black holes.
However since the equations of motion are highly nonlinear, it is rather hard to obtain
the exact analytic rotating black hole solutions. By introducing a small angular momentum as
a perturbation into a non-rotating system, Kim and Cai \cite{Kim:2007iw} studied slowly
Gauss-Bonnet rotating black hole solutions with one nonvanishing angular momentum.
It is worth to mention that the Lagrangian of
Gauss-Bonnet action only involves three terms constituted by the contracted
product of Ricci curvature and the Riemann curvature tensors. While, the resulting
field equations, obtained after variation with respect to the metric tensor, have seven terms.
If considering a higher order Lovelock gravity, the resulting field equation for third order
Lovelock gravity contains thirty-four terms \cite{Dehghani:2005zzb, Dehghani:2009zzb}.
Therefore, taking into account all
the relevant terms of the Lovelock action, obtaining slowly rotating black hole solutions
by solving the field equations in high dimensions is a very complicated
task. Note that the exact static Gauss-Bonnet black hole solutions
were obtained by working directly in the action in \cite{Boulware:1985wk, Cai:2003gr, Cai:2001dz}.
In this paper, we dedicate to investigate the slowly rotating black hole solutions
in Gauss-Bonnet gravity following the method. Apparently the lowest level
contribution of rotation should be proportional to $a^2$. Hence, linearly dependent on $a$,
the metric demonstrated in \cite{Kim:2007iw} is not applicable in this case. So, we need to find a
proper metric ansatz up to $a^2$.

This paper is organized as follows. In section 2, we present a new form metric and
obtain slowly rotating black hole solutions by working directly in the action. Section 3 is
devoted to a summary of the results.

\section{Slowly rotating Gauss-Bonnet black holes in five dimensions}  
\label{22s}
In this section, we analyze the slowly rotating Gauss-Bonnet black holes in five dimensional spacetimes.
In order to explore slowly rotating black holes by working directly in the action, the new metric
describing rotating black holes is expressed as
\begin{eqnarray}
ds^2=&-&\Big[f(r)+\frac{a^2}{l^2}+\frac{a^2(1-f(r))\cos^2\theta}{r^2}\Big]dt^2
+\Big[\frac{1}{f(r)}+\frac{a^2(\cos^2\theta f(r)-1)}{r^2f(r)^2}
-\frac{a^2}{l^2f(r)^2}\Big]dr^2\nonumber\\
&+&(r^2+a^2\cos^2\theta+\frac{a^2r^2}{l^2}\cos^2\theta)d\theta^2
+\Big[r^2+a^2+a^2\big(1-f(r)\big)\sin^2\theta\nonumber\\
&+&\frac{2a^2r^2\sin^2\theta}{l^2}\Big]\sin^2\theta d\phi^2
+2ar^2 p(r)\sin^2\theta dtd\phi+r^2\cos^2\theta d\varphi^2,\label{eq:1a}
\end{eqnarray}
where the parameter $a$ is a small quantity and the functions $f(r)$ and $p(r)$ are
two independent variables.

The action for Gauss-Bonnet theory with negative cosmological constant $\Lambda=-6/l^2$
in five dimensions is given by \cite{Cai:2001dz}
\begin{eqnarray}
{\cal I}&=&\frac{1}{16\pi G}\int d^{5} x\sqrt{-g}(-2\Lambda+R
+\alpha{\cal L}_2),\label{eq:2a}
\end{eqnarray}
where $\alpha$ is the Gauss-Bonnet coefficient with dimension $(length)^2$ and
is positive in the heterotic string theory. The second term $R$ is the Einstein-Hilbert
term and the third order term is the Gauss-Bonnet term
\begin{eqnarray}
{\cal L}_2=R_{\gamma\delta\lambda\sigma}R^{\gamma\delta\lambda\sigma}
-4R_{\gamma\delta}R^{\gamma\delta}+R^2.\label{eq:3a}
\end{eqnarray}
We also notice that the Lagrangian of Lovelock gravity is the sum of dimensionally
extended Euler densities \cite{Cai:2003kt}
\begin{eqnarray}
{\cal L}=\sum^{m}_{n=0}\alpha_n{\cal L}_n ,\label{eq:4a}
\end{eqnarray}
where $\alpha_n$ is an arbitrary constant and ${\cal L}_n$ is the Euler density of
a $2n$-dimensional manifold:
\begin{eqnarray}
{\cal L}_n=\frac{1}{2^n}\delta^{a_1b_1\cdots a_nb_n}_{c_1b_1\cdots c_nd_n}
R^{c_1d_1}_{~~~~a_1b_1}\cdots R^{c_nd_n}_{~~~~a_nb_n}.\label{eq:4b}
\end{eqnarray}
Here the generalized delta function is totally antisymmetric in both sets of indices.
Usually, we set ${\cal L}_0=1$ and hence $\alpha_0$ is just the cosmological constant.
${\cal L}_1$ gives us the usual Einstein-Hilbert term and ${\cal L}_2$ is the
Gauss-Bonnet term, and then it reads
\begin{eqnarray}
{\cal L}_2=\frac{1}{4}\delta^{a_1b_1a_2b_2}_{c_1b_1c_2d_2}
R^{c_1d_1}_{~~~~a_1b_1}R^{c_2d_2}_{~~~~a_2b_2}.\label{eq:5a}
\end{eqnarray}
In this paper, we focus on the Lagrangian of Gauss-Bonnet term with Eq.~(\ref{eq:5a}),
instead of the corresponding one in Eq.~(\ref{eq:3a}).

In principle, one can directly put the metric Eq.~(\ref{eq:1a}) into the action Eq.~(\ref{eq:2a}),
and derives out the equation of motion for functions $f(r)$ and $p(r)$. But, this will
become complicated, especially in higher order Lovelock gravity.
Fortunately, what we need is the term proportional to $a^2$, including the lower terms  in action.
According to the ansatz of metric, all non-vanishing components of Riemann tensors (up to $a^2$)
can be classified into three groups: (I) diagonal
form $R^{\hat{i}\hat{j}}_{~~\hat{i}\hat{j}}$, (II) off-diagonal form proportional to $a$ and
(III) off-diagonal form proportional to $a^2$. Some of key steps are given in appendixes.
For group (I), the elements read
\begin{eqnarray}
R^{12}_{~~12}&=&\bar{R}^{12}_{~~12}+\tilde{R}^{12}_{~~12},\quad
R^{1i}_{~~1i}=\bar{R}^{1i}_{~~1i}+\tilde{R}^{1i}_{~~1i},\nonumber\\
R^{2i}_{~~2i}&=&\bar{R}^{2i}_{~~2i}+\tilde{R}^{2i}_{~~2i},\quad
R^{ij}_{~~ij}=\bar{R}^{ij}_{~~ij}+\tilde{R}^{ij}_{~~ij}.\label{eq:6a}
\end{eqnarray}
where $3\leq i<j \leq 5$ and the Riemann tensors $\bar{R}^{12}_{~~12}$, $\bar{R}^{2i}_{~~2i}$
and $\bar{R}^{ij}_{~~ij}$ represent the components which are independent on parameter $a$, while
the tensors $\tilde{R}^{12}_{~~12}$, $\tilde{R}^{2i}_{~~2i}$ and $\tilde{R}^{ij}_{~~ij}$
are proportional to $a^2$. Denoted the Lagrangian of Gauss-Bonnet term from the contribution
of the case (I) by ${\cal L}_d$, it can be written as
\begin{eqnarray}
{\cal L}_d={\cal \bar{L}}_d+{\cal \tilde{L}}_d,\label{eq:7a}
\end{eqnarray}
where
\begin{eqnarray}
{\cal \bar{L}}_d&=&\frac{1}{4}\delta^{a_1b_1a_2b_2}_{c_1b_1c_2d_2}\bar{R}^{c_1d_1}_{~~~~a_1b_1}
\bar{R}^{c_2d_2}_{~~~~a_2b_2}\nonumber\\
&=&24\bar{R}^{12}_{~~12}\bar{R}^{34}_{~~34}+48\bar{R}^{13}_{~~13}\bar{R}^{24}_{~~24}
+24\bar{R}^{13}_{~~13}\bar{R}^{45}_{~~45}+24\bar{R}^{23}_{~~23}\bar{R}^{45}_{~~45},\nonumber\\
{\cal \tilde{L}}_d&=&\frac{1}{2}\delta^{a_1b_1a_2b_2}_{c_1b_1c_2d_2}\bar{R}^{c_1d_1}_{~~~~a_1b_1}
\tilde{R}^{c_2d_2}_{~~~~a_2b_2}\nonumber\\
&=&24\tilde{R}^{12}_{~~12}\times\bar{R}^{34}_{~~34}
+8\sum_{i=3}^5\tilde{R}^{1i}_{~~1i}\times(2\bar{R}^{24}_{~~24}+\bar{R}^{45}_{~~45})\nonumber\\
&+&8\sum_{j=3}^5\tilde{R}^{2j}_{~~2j}\times(2\bar{R}^{14}_{~~14}+\bar{R}^{45}_{~~45})\nonumber\\
&+&8(\tilde{R}^{34}_{~~34}+\tilde{R}^{35}_{~~35}
+\tilde{R}^{45}_{~~45})\times(\bar{R}^{12}_{~~12}+\bar{R}^{13}_{~~13}+\bar{R}^{23}_{~~23}).\label{eq:8a}
\end{eqnarray}

Unlike the static and spherically symmetric metric, there are some off-diagonal Riemann tensors.
With the help of  the properties of  Kronecker delta symbol, one can find that the contribution of
off-diagonal Riemann tensors in case (III) vanishes. The parts of Lagrangian from the off-diagonal
Riemann tensors in case (II) are obtained
\begin{eqnarray}
{\cal L}_{od}={\cal L}_{od1}+{\cal L}_{od2},\label{eq:9a}
\end{eqnarray}
where
\begin{eqnarray}
{\cal L}_{od1}&=&4(R^{12}_{~~34}R^{34}_{~~12}+R^{13}_{~~24}R^{24}_{~~13}+R^{14}_{~~23}R^{23}_{~~14}),
\nonumber\\
{\cal L}_{od2}&=&4(R^{12}_{~~24}R^{34}_{~~13}+R^{24}_{~~12}R^{13}_{~~34}+R^{12}_{~~24}R^{54}_{~~15}\nonumber\\
&+&R^{24}_{~~12}R^{15}_{~~54}+R^{13}_{~~34}R^{54}_{~~15}+R^{34}_{~~13}R^{15}_{~~54}).\nonumber
\end{eqnarray}
Furthermore, the Ricci scalar $R$ is given by
\begin{eqnarray}
R=\bar{R}+\tilde{R},\label{eq:10a}
\end{eqnarray}
where $\bar{R}$ is equal to $\frac{6(1-f(r)-rf(r)')}{r^2}-f(r)''$ and $\tilde{R}$
is proportional to $a^2$ and given in appendixes.

Varying the action Eq.~(\ref{eq:2a}) with regard to the function $p(r)$, we have
\begin{eqnarray}
0&=&[2\tilde{\alpha} f(r) r-2\tilde{\alpha} r-r^3]f(r)p(r)''
+[6\tilde{\alpha} f(r)-5r^2+2\tilde{\alpha} r f(r)'-6\tilde{\alpha}]f(r)p(r)'\nonumber\\
&+&3[-f(r)' r^2-2\tilde{\alpha} f(r)'+2r+2\tilde{\alpha} f(r)'f(r)
-2f(r)r+\frac{4r^3}{l^2}]p(r).\label{eq:11a}
\end{eqnarray}
Here we suppose that the coefficient of function $p(r)$ vanishes, and then
easily obtain
\begin{eqnarray}
f(r)=1+\frac{r^2}{2\tilde{\alpha}}\Big(1-\sqrt{1-\frac{4\tilde{\alpha}}{l^2}
+\frac{4\tilde{\alpha}m}{r^4}}\Big),\label{eq:12a}
\end{eqnarray}
where the Gauss-Bonnet coefficient $\alpha$ is rescaled to $\tilde{\alpha}/{2}$
and $m$ is a integral constant.
Hence, the Eq.~(\ref{eq:11a}) reduces to
\begin{eqnarray}
[2\tilde{\alpha} f(r) r-2\tilde{\alpha} r-r^3]p(r)''
+[6\tilde{\alpha} f(r)-5r^2+2\tilde{\alpha} r f(r)'-6\tilde{\alpha}]p(r)'=0.\label{eq:12b}
\end{eqnarray}
Then, the expression for function $p(r)$ can be written as
\begin{eqnarray}
p(r)=-\frac{C_2}{8\tilde{\alpha}m}\sqrt{1-\frac{4\tilde{\alpha}}{l^2}
+\frac{4\tilde{\alpha}m}{r^4}}+C_1,\label{eq:13a}
\end{eqnarray}
where the $C_1$ and $C_2$ are two integration constants. Let $C_2=4m$
and $C_1=\frac{1}{2\tilde{\alpha}}$, the Eq.~(\ref{eq:13a}) becomes
\begin{eqnarray}
p(r)=\frac{1}{2\tilde{\alpha}}\Big(1-\sqrt{1-\frac{4\tilde{\alpha}}{l^2}
+\frac{4\tilde{\alpha}m}{r^4}}\Big).\label{eq:14a}
\end{eqnarray}
In addition, we get another equation for functions $f(r)$ and $p(r)$ by varying the action
Eq.~(\ref{eq:2a}) respecting to $f(r)$. It is worth to point out that this equation identically
equals to zero when
$f(r)$ and $p(r)$ take the forms Eq.~(\ref{eq:12a}) and Eq.~(\ref{eq:14a}), respectively.
Notice that the expressions for functions $f(r)$ and $p(r)$ are identical
to the counterparts shown in \cite{Kim:2007iw}.

For the slowly rotating solution, the stationarity and rotational symmetry metric Eq.~(\ref{eq:1a})
admits two commuting Killing vector fields
$\xi_{(t)}=\frac{\partial}{\partial t}$ and $\xi_{\phi}=\frac{\partial}{\partial \phi}$.
The Killing vectors can be used to give a physical interpretation of the parameter $m$ and $a$.
Following the analysis given in \cite{Aliev:2007qi, Aliev:2006yk, Peng:2007pk, Zeng:2008zza},
one can obtain coordinate-independent definitions for these parameters. We have the integral
\begin{eqnarray}
M&=&-\frac{3}{32\pi G}\oint \xi^{\mu;\nu}_{(t)}d^3\Sigma_{\mu\nu},\nonumber\\
J&=&\frac{1}{16\pi G}\oint \xi^{\mu;\nu}_{(\phi)}d^3\Sigma_{\mu\nu},\label{eq:15a}
\end{eqnarray}
where the integral are taken over the three-sphere at spatial infinity,
\begin{eqnarray}
d^3\Sigma_{\mu\nu}=\frac{1}{3!}\sqrt{-g}\epsilon_{\mu\nu\alpha\beta\gamma}dx^{\alpha}\wedge
dx^{\beta}\wedge dx^{\gamma}.\label{eq:16a}
\end{eqnarray}
We can arrive at the mass $M$ and angular momentum $J$
\begin{eqnarray}
M=\frac{3m\Sigma_k}{16\pi G},\quad J=\frac{2Ma}{3}.\label{eq:17a}
\end{eqnarray}

\section{Concluding remarks}
\label{33s}
In this paper, we proposed an new metric and obtained the slowly rotating
Gauss-Bonnet black hole solutions in five dimensions by working directly in the action.
It is worth to note that the diagonal components of the metric also
involve $a^2$ besides the function $f(r)$. Moreover, $g_{t\phi}$ is proportional to $r^2p(r)$
as to make the equation for $p(r)$ much simple. By discarding any terms involving $a^3$
and higher powers in $a$ in the action, we got the exact form for function $p(r)$, while the
function $f(r)$ still kept the form of the static solutions. In addition, we described the
Killing isometries of the metric and presented
its mass parameter and angular momentum of the black holes.

The advantage of this method is avoiding the equations of motion to arrive
at the slowly rotating black holes solutions. Although we only considered the
slowly rotating black holes in five dimensional spacetimes,
this method is still valid for general Gauss-Bonnet gravity in higher dimension
including charge. Furthermore, in general Lovelock gravity, the Einstein
equation  must involve the terms derived from the action of Lovelock gravity,
which  becomes very hard if $n>3$.
But, in present case, the key is to find the action, which is
possible to carry out for slowly rotating and charge cases. Besides, this approach can
also be used to find the slowly rotating solutions with involving multiple angular
momenta in different orthogonal planes of rotation. We will discuss them elsewhere.

\section{Appendixes}
\indent
From the metric Eq.~(\ref{eq:1a}), we obtain some of the intermediate steps of the calculation.

\textbf{Riemann tensors}. The non-vanishing Riemann tensors can be classified into three
groups: (I) diagonal form $R^{\hat{i}\hat{j}}_{~~\hat{i}\hat{j}}$, (II) off-diagonal form
proportional to $a$ and (III) off-diagonal form proportional to $a^2$.

group (I):
\begin{eqnarray}
R^{12}_{~~12}&=&\bar{R}^{12}_{~~12}+\tilde{R}^{12}_{~~12},\quad
\bar{R}^{12}_{~~12}=-\frac{f(r)''}{2},\nonumber\\
\tilde{R}^{12}_{~~12}&=&a^2\left\{[(-1+\frac{r f(r)'+r^2 f(r)''}{2f(r)}
-\frac{r^2f(r)'^2}{4f(r)^2})p(r)^2
+(\frac{r^2f(r)'p(r)'}{f(r)}-5rp(r)'\right.\nonumber\\
&-&\left.r^2p(r)'')\frac{p(r)}{2}
-\frac{r^2p(r)'^2}{4}+\frac{f(r)'}{2r^3f(r)}+\frac{f(r)'^2}{4r^2f(r)^2}
-\frac{f(r)''}{2r^2f(r)}]\sin^2\theta\right.\nonumber\\
&+&\left.(\frac{-2f(r)'}{r^3}+\frac{f(r)''}{2r^2}+\frac{3f(r)-3}{r^4})\cos^2\theta\right\},\nonumber
\end{eqnarray}
\begin{eqnarray}
R^{1i}_{~~1i}&=&\bar{R}^{1i}_{~~1i}+\tilde{R}^{1i}_{~~1i},\quad
\bar{R}^{1i}_{~~1i}=-\frac{f(r)'}{2r}, \quad (3\leq i\leq 5) ,\nonumber\\
\tilde{R}^{13}_{~~13}&=&\left\{-\frac{p(r)^2}{f(r)}+\frac{1}{f(r)r^4}+\frac{f(r)'}{r^3}-\frac{f(r)}{r^4}
+[-\frac{r p(r)^2f(r)'}{2f(r)}+\frac{r p(r)p(r)'}{2}\right.\nonumber\\
&-&\left.\frac{f(r)+1}{r^4}+p(r)^2-\frac{f(r)'}{2f(r)r^3}+\frac{2}{r^4f(r)}
-\frac{2p(r)^2}{f(r)}+\frac{f(r)'}{r^3}]\sin^2\theta\right\}a^2.\nonumber
\end{eqnarray}
\begin{eqnarray}
\tilde{R}^{14}_{~~14}&=&\left\{[-p(r)^2+\frac{f(r)+f(r)'^2}{r^4}-r p(r)'p(r)-\frac{r^2p(r)'^2}{4}
-\frac{f(r)'}{2r^3f(r)}\right.\nonumber\\
&-&\left.\frac{1}{f(r)r^4}+\frac{p(r)^2}{f(r)}+\frac{f(r)'}{4r^2}
-\frac{f(r)'f(r)}{2r^3}+\frac{r f(r)'p(r)^2}{2f(r)}]\sin^2\theta\right.\nonumber\\
&-&\left.\frac{p(r)^2}{f(r)}+\frac{1}{f(r)r^4}+\frac{f(r)'}{r^3}-\frac{f(r)}{r^4}\right\}a^2\nonumber\\
\tilde{R}^{15}_{~~15}&=&\left\{[\frac{2p(r)^2+r f(r)'p(r)^2}{f(r)}-p(r)^2-\frac{f(r)'}{2r^3f(r)'}
-\frac{1}{r^4f(r)}-\frac{p(r)'p(r)}{2}]\sin^2\theta\right. \nonumber\\
&+&\left. [\frac{f(r)'}{2r^3}-\frac{f(r)}{r^4}]\cos^2\theta+\frac{1}{r^4}\right\}a^2.
\end{eqnarray}
\begin{eqnarray}
R^{2i}_{~~2i}&=&\bar{R}^{2i}_{~~2i}+\tilde{R}^{2i}_{~~2i},\quad
\bar{R}^{2i}_{~~2i}=-\frac{f(r)}{2r},\quad (3\leq i \leq 5),\nonumber\\
\tilde{R}^{23}_{~~23}&=&\tilde{R}^{25}_{~~25}=a^2[\frac{f(r)'}{r^3}+\frac{2(1-f(r))}{r^4}]\cos^2\theta, \nonumber\\
\tilde{R}^{24}_{~~24}&=&\left\{[\frac{f(r)'^2+2f(r)''f(r)}{4r^2}-1-\frac{r^2p(r)p(r)''}{2}
-\frac{3f(r)f(r)'}{2r^3}+\frac{f(r)^2}{r^4}\right.\nonumber\\
&-&\left.\frac{r^2p(r)'^2}{4}-\frac{3rp(r)p(r)'}{2}]\sin^2\theta
+4(2-2f(r)+f(r)'r)\right\}a^2.
\end{eqnarray}
\begin{eqnarray}
R^{ij}_{~~ij}&=&\bar{R}^{ij}_{~~ij}+\tilde{R}^{ij}_{~~ij},\quad
\bar{R}^{ij}_{~~ij}=\frac{1-f(r)}{r^2}, \quad (3\leq i<j \leq 5),\nonumber\\
\tilde{R}^{34}_{~~34}&=&\left\{[\frac{6-5f(r)-f(r)^2}{r^4}+\frac{f(r)f(r)'}{2r^3}+\frac{10}{r^2}
-\frac{r p(r)p(r)'}{2}]\sin^2\theta\right.\nonumber\\
&-&\left.\frac{9}{r^2l^2}-\frac{6(1-f(r))}{r^4}\right\}a^2, \nonumber\\
\tilde{R}^{35}_{~~35}&=&2a^2\cos^2\theta(-\frac{1}{r^2l^2}+\frac{f(r)-1}{r^4}), \nonumber\\
\tilde{R}^{45}_{~~45}&=&\left\{[\frac{f(r)f(r)'}{2r^3}-\frac{r p(r)p(r)'}{2}
+\frac{3}{r^2l^2}+\frac{2-f(r)^2-f(r)}{r^4}]\sin^2\theta\right.\nonumber\\
&+&\left.\frac{2(f(r)-1)}{r^4}-\frac{2}{r^2l^2}\right\}a^2.
\end{eqnarray}

group (II):
\begin{eqnarray}
R^{12}_{~~34}&=&-ar^2\sin\theta\cos\theta p(r)',\quad
R^{13}_{~~24}=-\frac{a\cos\theta\sin\theta p(r)'}{2f(r)},\nonumber\\
R^{14}_{~~32}&=&-\frac{a\cos\theta p(r)'}{2f(r)\sin\theta},\quad
R^{12}_{~~24}=-\frac{ar\sin^2\theta}{2}(3p(r)'+r p(r)''),\nonumber\\
R^{13}_{~~34}&=&R^{15}_{~~54}=-\frac{ar\sin^2\theta}{2}p(r)'.
\end{eqnarray}

group (III):
\begin{eqnarray}
R^{12}_{~~13}&=&[-\frac{f(r)'}{2f(r)r^2}-p(r)^2 r+\frac{p(r)^2f(r)'r^2}{2f(r)}
-p(r)p(r)'r^2+\frac{3f(r)-3}{r^3}\nonumber\\
&-&\frac{f(r)'}{r^2}]a^2\sin\theta\cos\theta,\nonumber\\
R^{24}_{~~34}&=&[-p(r)p(r)'r^2+\frac{3f(r)(1-f(r))}{r^3}+\frac{3f(r)'f(r)}{2r^2}]a^2\sin\theta\cos\theta.
\end{eqnarray}

With regard to Ricci scalar, the term $\tilde{R}$ which is proportional to $a^2$ is given by
\begin{eqnarray}
\tilde{R}&=&\left\{[(\frac{r^2f(r)''}{f(r)}-8-\frac{r^2f(r)'^2}{2r^2f(r)^2}+\frac{4rf(r)'}{f(r)}
+\frac{8}{f(r)})p(r)^2+(14r-\frac{r^2f(r)'}{f(r)})p(r)'\right. \nonumber\\
&-&\left. \frac{3r^2p(r)'^2}{2}-\frac{2f(r)'}{f(r)r^3}-\frac{2f(r)f(r)'}{r^3}+\frac{f(r)'^2}{r^2}
-\frac{8}{r^4f(r)}+\frac{f(r)f(r)''}{r^2}-\frac{2f(r)'}{r^3}\right. \nonumber\\
&+&\left.\frac{f(r)'^2}{2f(r)^2r^2}-\frac{f(r)''}{r^2}-\frac{2f(r)^2}{r^4}
+\frac{20}{r^4}-\frac{f(r)''}{f(r)r^2}-\frac{10f(r)}{r^4}]\sin^2\theta
\frac{4p(r)^2}{f(r)}\right. \nonumber\\
&-&\left.\frac{14}{r^4}+ \frac{4}{f(r)r^4}+\frac{6f(r)'}{r^3}+\frac{f(r)''}{r^2}
+\frac{10f(r)}{r^4}+\frac{30\sin^2\theta-26}{r^2l^2}\right\}a^2.
\end{eqnarray}

{\bf Acknowledgment }
This work has been supported by the Natural Science Foundation of China
under grant Nos.10875060, 10975180 and 11047025.


\end{document}